\documentclass[superscriptaddress,twocolumn,prl,showpacs,floatfix]{revtex4}
\usepackage[dvips]{graphicx}
\usepackage{amsmath,amsfonts,amssymb,bm,ulem}
\usepackage{graphicx}

\newcommand{\e}{\text{eff}}

\begin{document}
%
%
\title{Nonlinear high-frequency hopping conduction in two-dimensional
arrays of Ge-in-Si quantum dots: Acoustic methods}

\author{I.~L.~Drichko}
\affiliation{A. F. Ioffe Physico-Technical Institute of Russian
Academy of Sciences, 194021 St. Petersburg, Russia}
\author{A.~M.~Diakonov}
\affiliation{A. F. Ioffe  Physico-Technical Institute of Russian
Academy of Sciences, 194021 St. Petersburg, Russia}
\author{V.~A.~Malysh}
\affiliation{A. F. Ioffe  Physico-Technical Institute of Russian
Academy of Sciences, 194021 St. Petersburg, Russia}
\author{I.~Yu.~Smirnov}
\affiliation{A. F. Ioffe  Physico-Technical Institute of Russian
Academy of Sciences, 194021 St. Petersburg, Russia}
\author{E.~S.~Koptev}
\affiliation{Institute of Semiconductor Physics, Siberian Branch of
Russian Academy of Sciences, 630090 Novosibirsk, Russia}
\author{A.~I.~Nikiforov}
\affiliation{Institute of Semiconductor Physics, Siberian Branch of
Russian Academy of Sciences, 630090 Novosibirsk, Russia}
\author{N.~P.~Stepina}
\affiliation{Institute of Semiconductor Physics, Siberian Branch of
Russian Academy of Sciences, 630090 Novosibirsk, Russia}
\author{Y.~M.~Galperin}
\affiliation{Department of Physics, University of Oslo, PO Box 1048
Blindern, 0316 Oslo, Norway} \affiliation{Centre for Advanced Study,
Drammensveien 78, 0271 Oslo, Norway} \affiliation{A. F. Ioffe
Physico-Technical Institute of Russian Academy of Sciences, 194021
St. Petersburg, Russia}
\author{J.~Bergli}
\affiliation{Department of Physics, University of Oslo, PO Box 1048
Blindern, 0316 Oslo, Norway}
\date{\today}
\begin{abstract}
Using
acoustic methods we have measured nonlinear AC conductance in 2D arrays of Ge-in-Si quantum dots.
 The combination of experimental results and modeling of AC conductance of  a dense lattice of localized states
leads us to the conclusion that the main mechanism of AC conduction
in hopping systems with large localization length is due to the
charge transfer within large clusters, while the main mechanism
behind its non-Ohmic behavior is charge heating by absorbed power.
\end{abstract}
\maketitle

 \section{Introduction} \label{introduction}

Nonlinear hopping conduction in various materials and devices was
extensively
studied, see, e.g., \cite{Gershenson2000,Ovadyahu2011} and references
therein.
%
Despite relatively large amount of experimental and
theoretical work aimed at nonlinear phenomena in static (DC) hopping conduction,
the understanding in this research area is far from being complete. The
common knowledge
is that nonlinear effects in hopping conduction
are determined by interplay between the so-called field effects
(field-induced deformation of percolation paths governing the
conduction) and heating of the charge carriers.
The
non-Ohmic effects are
usually expressed in terms of the
dimensionless ratio $eE\mathcal{L}/kT$, where $e$ is the electron
charge, $E$ is the electrical field strength, $T$ is the
temperature, $k$ is the Boltzmann constant, while $\mathcal L$ is
some characteristic length.
Theoretical considerations
\cite{theory}
based on different models
lead to apparently similar predictions: non-Ohmic effects become noticeable at $eE\mathcal{L}/kT \gtrsim 1$. However, predictions of both
magnitude $\mathcal{L}$ and its temperature dependence significantly
differ.
%
Interplay between the field and the heating effects in DC non-Ohmic hopping conduction
is essentially dependent on the relation between the localization length $\xi$
and a typical distance between electrons~\cite{Gershenson2000}.
In two-dimensional (2D)
systems with large localization length the hopping regime is
qualitatively different in many respects from the conventional hopping
in systems with a small $\xi$~\cite{ES1984}. In particular, the
nonlinear effects in the 2D hopping transport in systems with a large $\xi$ are
caused by \textit{electron heating}~\cite{Gershenson2000}.

Much less attention has been paid to non-Ohmic effects caused by a
high-frequency (AC) field. Most studies have been done on
modifications of DC conductance under electromagnetic radiation.
They revealed a rather complicated picture of nonlinear behaviors
(see \cite{Ovadyahu2011} and references therein), which requires
more theoretical effort for complete understanding.

High-frequency hopping conductance is conventionally analyzed within
the framework of the so-called \textit{two-site model}, according to
which an electron hops between states with close energies localized
at two different centers. These states form pair complexes, which do
not overlap. Therefore, they do not contribute to the DC conduction,
but are important for the AC response. Being very simple, the
two-site model has been extensively studied, see for a review
\cite{Efros1985,Efros1985a,Galperin1991} and references therein. As
is well known \cite{Efros1985a}, there are two specific
contributions to the high-frequency absorption. The first
contribution, the so-called resonant, is due to direct absorption of
microwave quanta accompanied by interlevel transitions. The second
one, the so-called relaxation, or phonon assisted, is due to
phonon-assisted transitions, which lead to a lag of the levels
populations with respect to the microwave-induced variation in the
interlevel spacing. The relative importance of the two mechanisms
depends on the frequency $\omega$, the temperature $T$, as well as
on sample parameters. The most important of them is the relaxation
rate $\gamma (T)$ of symmetric pairs with interlevel spacing
$\sim$$kT$. At $\omega \lesssim \sqrt{\gamma kT/\hbar}$ the
relaxation contribution to the real part of AC conductivity
$\sigma_1 (\omega)$ dominates. In this case the imaginary component
of AC condutivity $\sigma_2 \gtrsim \sigma_1$, and $\sigma_1
(\omega) \propto \omega$ with logarithmic accuracy. According to the
two-site model, the non-Ohmic conductance \textit{decreases} with
the field amplitude \cite{Galperin1991,Galperin1997}.

In this work we 
study 
 nonlinear AC
conductance in the samples containing dense arrays of Ge-in-Si quantum
dots (QDs) using probe-free acoustic method~\cite{Drichko2000}.
We will show that AC conduction of dense QD arrays is
similar to
that in  hopping semiconductors with large localization length.

\section{Experiment} \label{experiment}

\begin{figure}[h]
\centerline{
\includegraphics[width=1\columnwidth]{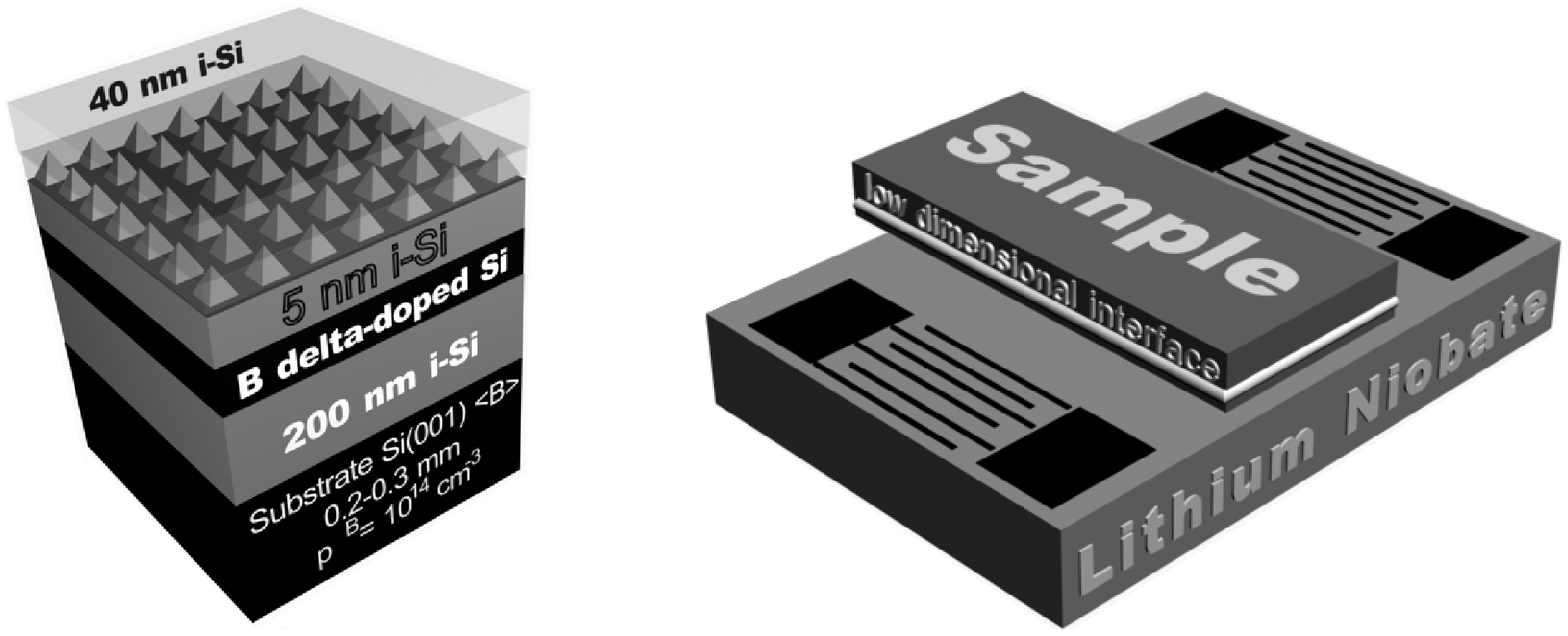} }
\caption{Sketch of
the sample (left) and experimental setup (right). \label{fig1}}
\end{figure}

\paragraph{Samples} We study B-doped arrays of Ge-in-Si QDs with
densities $n=(3-4)\times 10^{11}$ cm$^{-2}$ and filling factors $\nu
\approx 2.5$ - 2.85~\cite{Stepina2010}. The samples were grown by
Stranski-Krastanov molecular beam epitaxy on a (001) Si substrate.
The QD array layer was located at 40 nm from the sample surface, see
the sketch in the left panel of Fig.~\ref{fig1}. QDs are shaped as
pyramids with a height of 10-15 {\AA} and a square base having the
side of 100-150 {\AA}. The samples were $\delta$-doped with B, the
density in the doped layer being $\sim 10^{12}$ cm$^{-2}$. In these
systems two lowest states in Ge QDs are occupied by holes, and the
third level is partly occupied. The 4th state is split from the 3rd
one by 18-23 meV \cite{Yakimov2001}. Therefore, the 4th level is
expected to be empty in the temperature domain of interest.

We studied AC conductance of two samples, which were annealed and
were close in conductance. The aim of annealing was increasing of
the sample conductance; during the annealing the dots spread and the
distance between them decreases from initial $\sim$15 nm to almost
overlapping. The samples studied in \cite{Drichko2005}  had larger
conductance and therefore AC hopping conductance was not observed
there.
%

\paragraph{Procedure}
We used the probeless acoustic method to determine
the complex ac conductance,
$\sigma \equiv \sigma_1-i\sigma_2$,
from attenuation, $\Gamma$, and velocity, $V$, of a
surface acoustic wave (SAW) induced on the surface of a LiNbO$_3$
piezoelectric crystal by interdigital transducers, see right panel of Fig.~\ref{fig1}. The sample is pressed
to the crystal surface by a spring. A SAW-induced  AC electric field
penetrates  into the sample and interacts with charge carriers (holes).
As a result, both $\Gamma$ and $V$ acquire additional contributions, which can be
related to complex $\sigma$~\cite{Drichko2000}.
%
%
We single out these contributions by measuring
the changes induced
by external transverse magnetic field, $\Delta \Gamma (B) \equiv \Gamma
(0)-\Gamma (B)$, and similarly $\Delta V (B)$. For these particular
samples $\Delta V$ is not measurable thus indicating that $\sigma_2 \ll
\sigma_1$. Therefore, further we will discuss only $\Delta \Gamma$ and
$\sigma_1$. Since the hole absorption vanishes in very high magnetic
fields we extrapolate experimentally found $\Delta \Gamma (B)$ to $B$$\to$$\infty$
 and in this way find $\Gamma (0) \equiv \Gamma$ versus temperature and SAW
frequency at frequencies 30-414 MHz in the temperature domain 1.8-13 K.

\begin{figure}[t]
 \centerline{
\includegraphics[width=1\columnwidth]{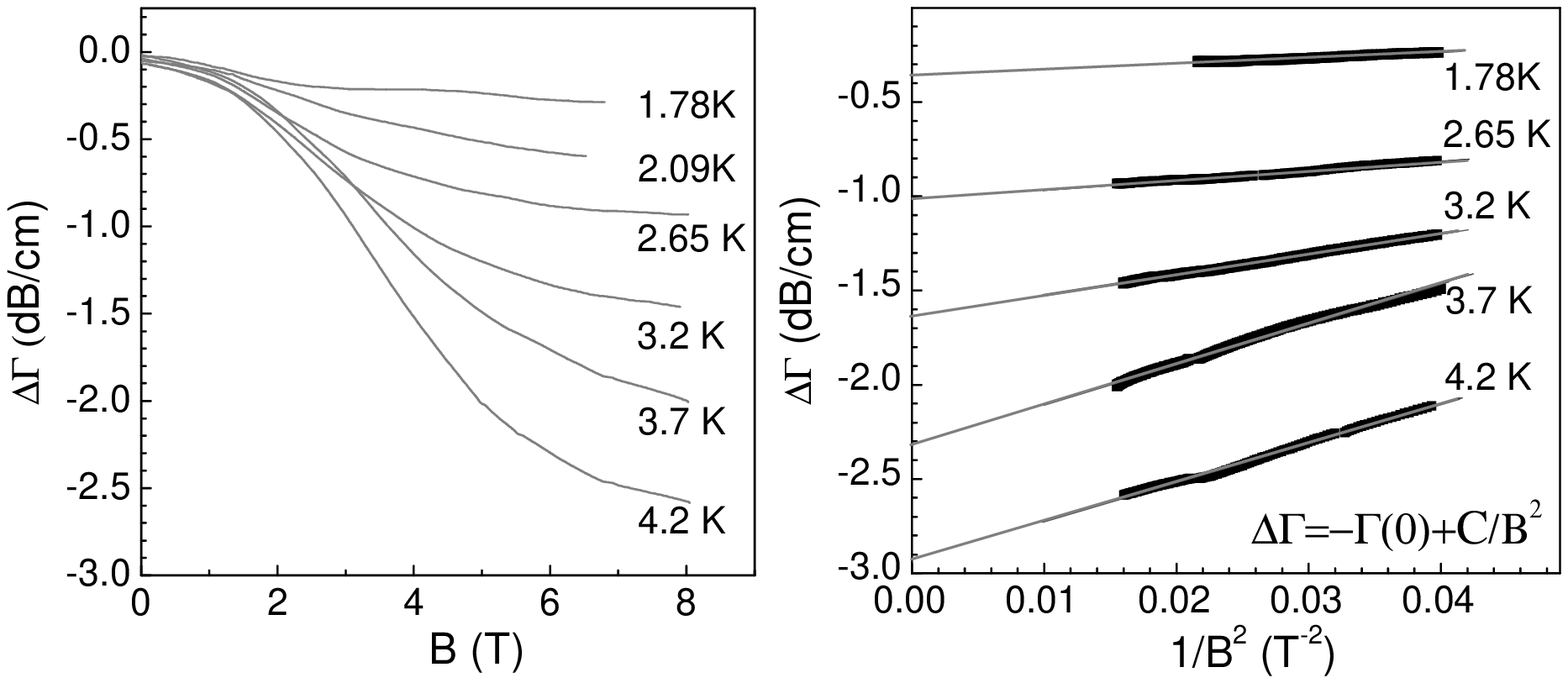} }
 \caption{Left --
Magnetic field dependence of $\Delta \Gamma$ at frequency 143 MHz for different
temperatures shown at the curves. Right --The same dependences replotted
as function of $B^{-2}$. Sample 1.\label{fig2}}
 \end{figure}

\paragraph{Linear regime}
Shown in the left panel of Fig.~\ref{fig2} are
magnetic field dependences of $\Delta \Gamma $ at $f= 143$ MHz and $B
\le 8$ T for different temperatures.
These dependences
are replotted in the right panel as functions of $B^{-2}$, which can be
represented as straight lines. The attenuation coefficient at $B=0$,
$\Gamma$,
is determined by
crossing of these lines with the axis $B^{-2}=0$.
%
%
For the situation relevant to present experiment the
general  expression~\cite{Drichko2000} relating $\Gamma$ with $\sigma_1 (\omega)$
can be simplified as
\begin{equation} \label{001}
 \Gamma \, (\textrm{dB}/\textrm{cm})=4.34 K^2
qA(q)\, 4\pi t(q)\, \sigma_1 (\omega)/\varepsilon_s V \, .
\end{equation}
Here $q \equiv \omega/V$ is the SAW wave vector, $K$ is
the piezoelectric coupling constant in the LiNbO$_3$, $A(q)$ and $t(q)$
are geometric factors given in \cite{Drichko2000}, which depend on the
dielectric constants of the sample ($\varepsilon_s$), vacuum
($\varepsilon_0$) and substrate ($\varepsilon_1$), as well as on the depth
of the QD layer and the clearance $a$ between the sample and
substrate. The above expression is valid provided
$ \sigma_2  \ll \sigma_1\ll \varepsilon_s
V/4\pi t(q)$
that is the case in the present
situation.

\begin{figure}[h]
\centerline{
\includegraphics[width=1\columnwidth]{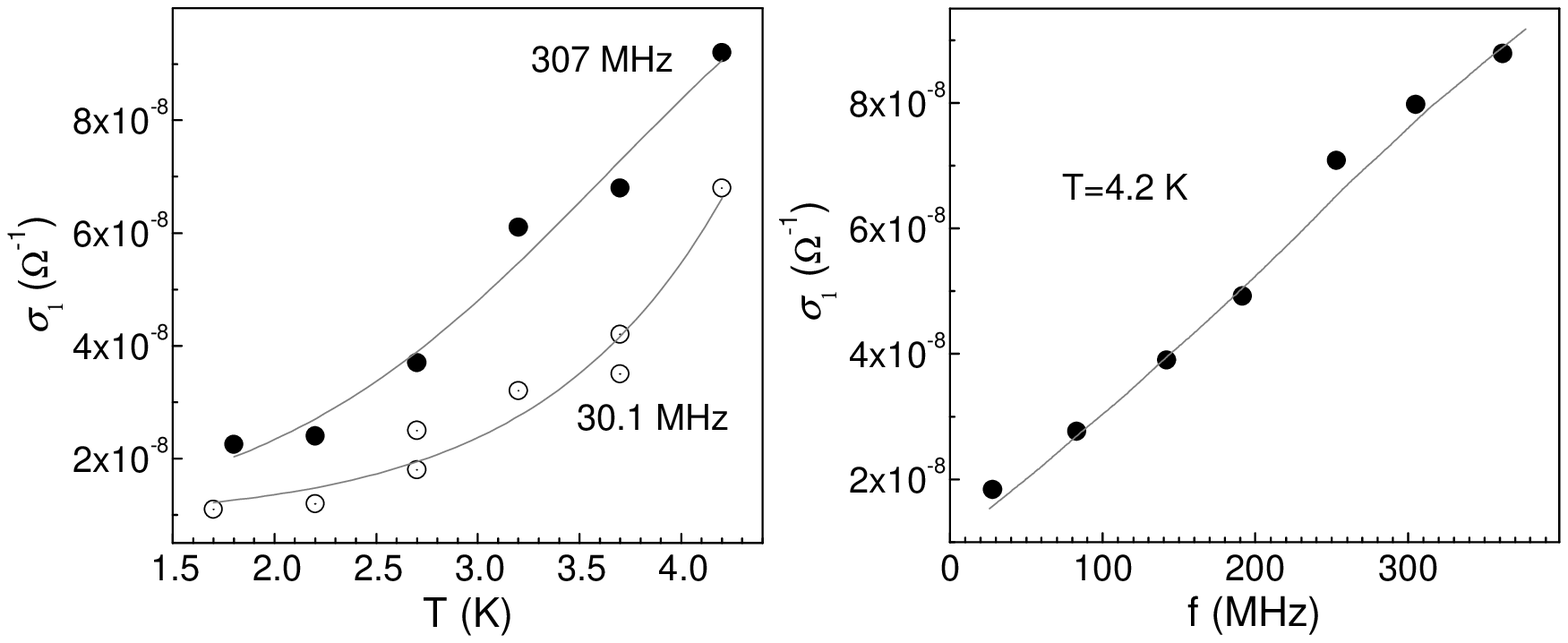}
}
 \caption{Left -
Temperature dependences of $\sigma_1(\omega)$ in the sample 1 for
$f=30.1$ and 307 MHz. $a=5\times 10^{-5}$ cm. Right -- Frequency
dependence of $\sigma_1$ in the sample 2 at $T=4.2$~K.
$a=4\times 10^{-5}$ cm. The lines are guides to the eye.\label{fig3}}
 \end{figure}

Shown in the left panel of Fig.~\ref{fig3} are temperature dependences
of $\sigma_1$ for the sample 1 at frequencies 30.1 and
307 MHz found from Eq.~(\ref{001}). The clearance $a$ was
determined
from Newton's rings.
 Frequency dependence
of $\Gamma$ for the sample 2 at $T=4.2$~K is shown in the
right panel.

\begin{figure}[b]
\centerline{ \includegraphics[width=1\columnwidth]{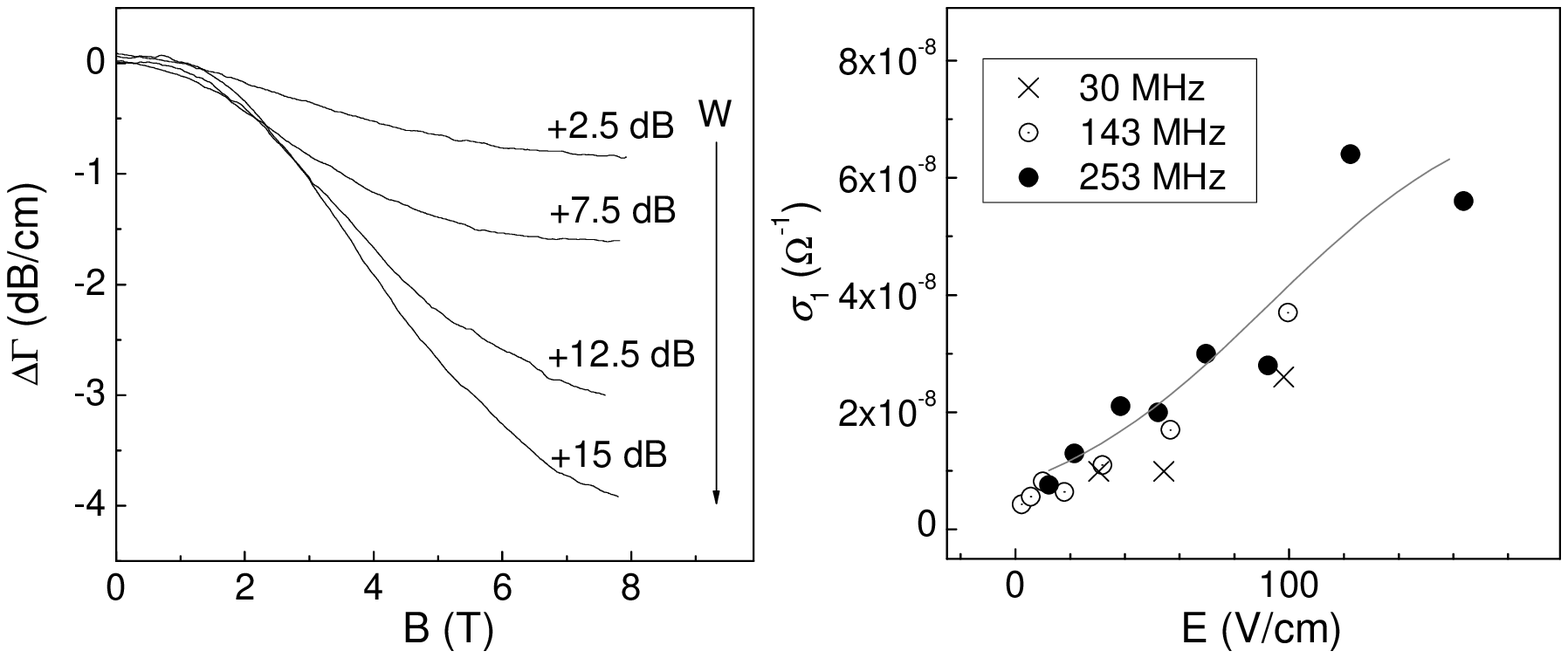}
 } \caption{ Left --
Magnetic field dependences of $\Delta \Gamma$ at frequency $f=253$ MHz
at $T=1.8$ K for different input SAW intensities. Right -- The
dependence $\sigma_1 (E)$ for $T=1.8$ K  with the line guided to the eye. Sample 1.\label{fig4}}
\end{figure}

\paragraph{Nonlinear regime} Shown in the left panel of Fig.~\ref{fig4}
are magnetic field dependences of $\Delta \Gamma$ at frequency $f=253$
MHz at $T=1.8$ K for different input SAW intensities.
Similarly to the linear regime, the values $\Gamma (0)$, and $\sigma_1$ were determined
by extrapolation to $B$$\to$$\infty$. The intensities $W$ were determined
as follows. Firstly, the signal having passed through the system was
compared with the input signal from the generator. This comparison
allowed determining the transformation losses in both transducers plus
the losses in input and output transmission lines (which are considered
to be equal). The intensity was changed by synchronous changing of the
calibrated attenuators installed after the generator and after the
output transducer keeping the summary attenuation (in dB) constant.
Knowing the transformation losses in one transducer and in the input
line we find the acoustic intensity after the input transducer. Similar
measurements were done for $f=30$, 143 and 253 MHz.

Since for each frequency the inequality $\Gamma L \ll 1$ (where
$\Gamma$ is the attenuation coefficient in cm$^{-1}$ while $L=4-5$
mm is the sample length) is met, one can neglect the difference
between the input and output intensities while calculating the
typical amplitude $E$ of the SAW-induced electrical field. It is
evaluated from the energy conservation law $2 \Gamma W = \sigma_1(E)
E^2/2$. This procedure is valid since due to the inequality $\Gamma
L \ll 1$ higher harmonics of the signal are weak. The obtained in
this way dependence $\sigma_1 (E)$ for sample 1, at $T=1.7$ K and
different frequencies is shown on the right panel of
Fig.~\ref{fig4}. One observe very weak frequency dependence.

The regime of weak non-Ohmic behavior ($E \le 100$ V/cm) was
investigated in more detail for the sample 2 at $f=142$ MHz and
different temperatures, $T=3$, 3.3, 4.2 and 7 K.
\begin{figure}[t]
\centerline{ \includegraphics[width=.6\columnwidth]{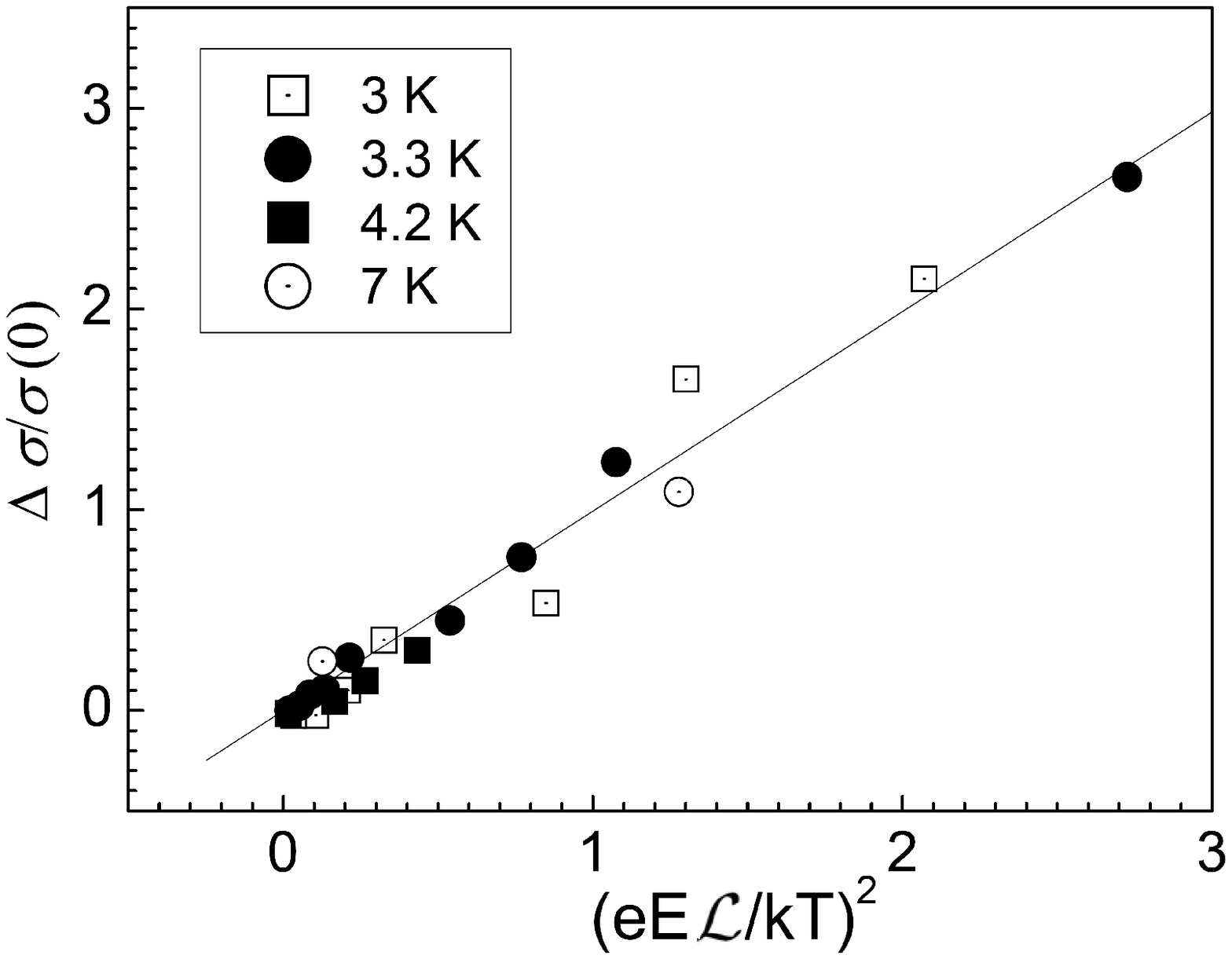}
 }
 \caption{
Dependence of $\Delta \sigma/\sigma_1(0)$
on the dimensionless parameter $(eE
\mathcal{L}/kT)^2$.
Here $\Delta \sigma (E) \equiv \sigma_1(E)- \sigma_1(0)$, while
$\mathcal{L}$ is a characteristic length chosen
as $7.5 \times10^{-6}$ cm. $f=142$ MHz , $T=3$, 3.3, 4.2 and 7 K. Sample 2. The line is guide to the eye.
\label{fig6}}
\end{figure}
The dependences $\Delta \sigma/\sigma_1(0) \equiv [
\sigma_1(E)/\sigma_1(0)-1]$ on dimensionless parameter $(eE
\mathcal{L}/kT)^2$ are shown in Fig.~\ref{fig6}.
The characteristic length $\mathcal{L}$ is
 chosen as $7.5 \times10^{-6}$ cm in order to have
 $\Delta \sigma/\sigma_1(0) =1$ at  $(eE \mathcal{L}/kT)^2=1$.
 One observes an approximate data collapse.

\section{Simulations}

Since we are not aware of any analytical theory for AC hopping
conduction in dense QD arrays we have performed numerical
simulations for a model system consisting of a $L\times L$ square
lattice of randomly occupied localized states~\cite{davies}.
We put  $L=100$
having in mind that for this size we know that finite size effects are not
appreciable in DC conductance~\cite{aurora}. Each of the states can be
either empty or occupied (double occupancy is not allowed). The simulations were
performed for the filling factor  of 1/2, so that
the number of electrons is half the number of sites. Disorder is
introduced by assigning a random energy in the range $[-U,U]$ to each
site (in our numerics we have used $U=e^2/d$ where $d$ is the lattice constant).
 Each site is also given a
compensating charge $\nu e$ so that the overall system is charge
neutral. The charges interact via the Coulomb interaction. In the
following
 the
unit of energy is chosen as the Coulomb energy of unit charges on
nearest neighbor sites, $e^2/d$. The unit of temperature is then $e^2/dk$.

To simulate the time evolution we used the dynamic Monte Carlo method
introduced in Ref.~\cite{tsigankov} for simulating DC transport.
The only difference is that we apply an AC electric field
$E=E_0\cos\omega t$ and that, following Ref.~\cite{tenelsen},
we use for the transition rate of an electron from site $i$ to site $j$ the
formula
\begin{equation}
\gamma_{ij} = {\tau_0}^{-1}e^{-2 r_{ij}/\xi}\min
  \left(e^{- \varepsilon_{ij}/kT},1\right) \label{eq:01}
\end{equation}
where $\varepsilon_{ij}$ is the energy of the phonon and $r_{ij}$ is
the distance between the sites.  $\tau_0$ contains material
dependent and energy dependent factors, which we approximate by
their average value; we consider it as constant and its value, of
the order of $10^{-12}$ s, is chosen as our unit of time.
Consequently, $\omega$ is measured in units of $\tau_0^{-1}$
while the electric field is measured in units of $e/d^2$ .
The conductance
 (in Siemens) is expressed in units of
$4\pi \epsilon_0 d/\tau_0$.
Note that our model
is oversimplified comparing to realistic dense QD arrays. In particular, the model
assumes a hydrogen-like wave function with
localization length $\xi$ and independent of the phonon wave vector
electron-phonon coupling, we ignore dielectric susceptibility of the host, etc.
Therefore, we hope to reproduce the behavior of AC conductance only qualitatively.
However, we believe that the model reproduces main physics of the dense arrays --
interplay between Coulomb correlations and disorder as well as relatively large ratio between $\xi$ and a distance between the dots.
To decrease the number of independent parameters we put $\xi = d$.

\begin{figure}[t]
\begin{center}
\includegraphics[width=0.49\columnwidth]{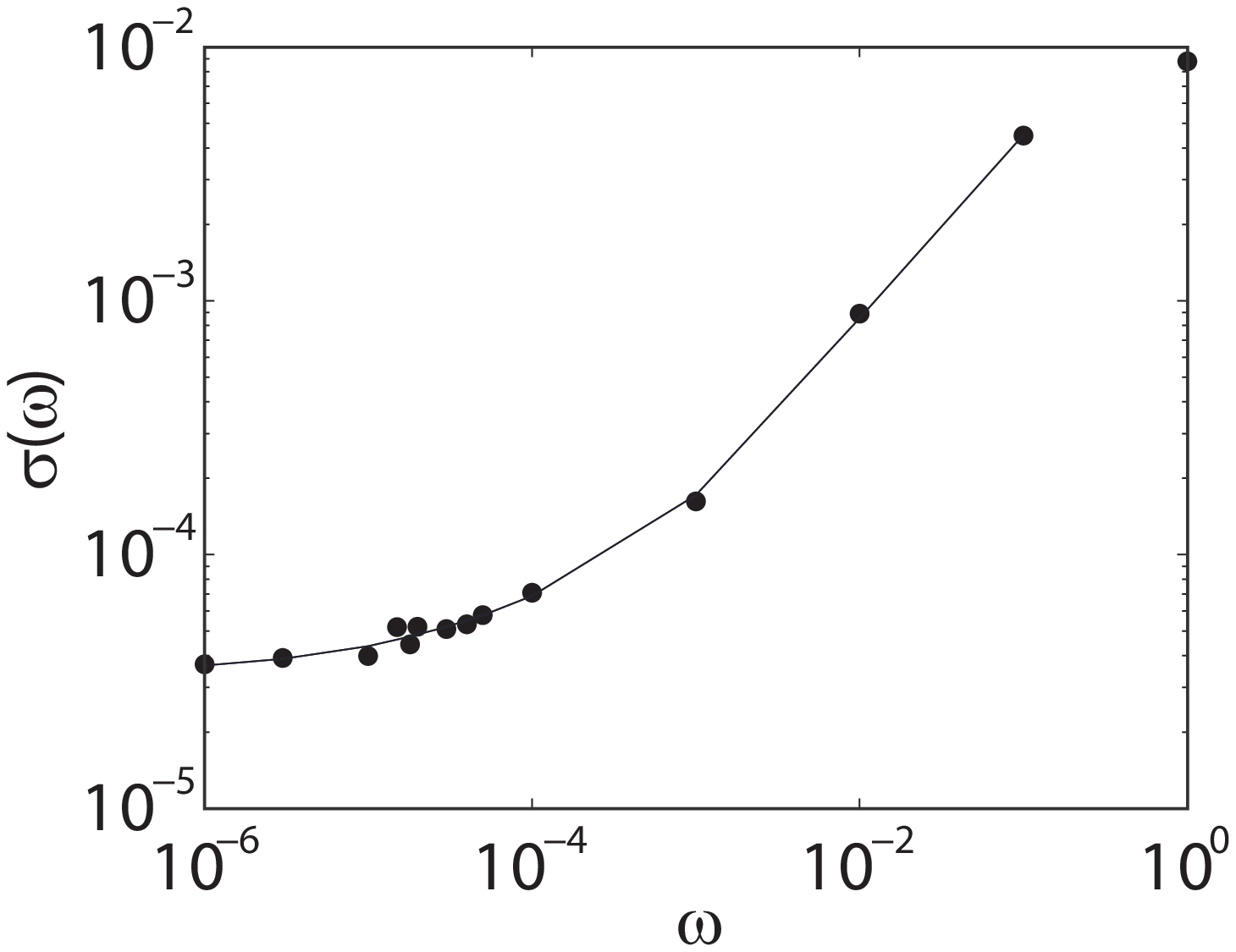} \hfill \includegraphics[width=0.49\columnwidth]{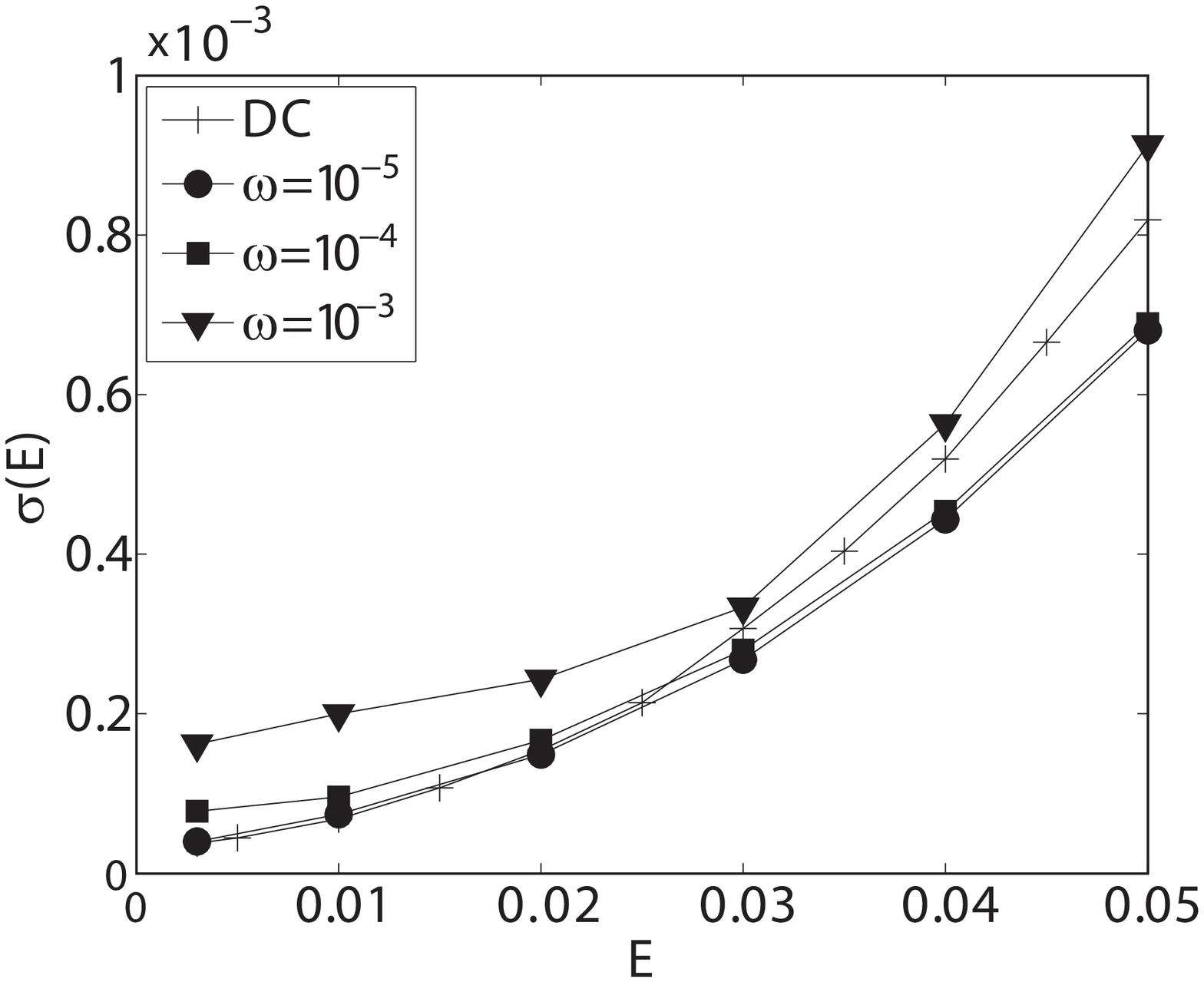}
\caption{\label{condAC}AC conductivity as function of frequency (left panel) and of electrical field (right panel)  for
$T=0.03$.
}
\end{center}
\end{figure}

Shown in Fig.~\ref{condAC} (left panel) is  the conductance versus
frequency  for $T=0.03$ (corresponding to the Efros-Shklovskii regime) and
 $E_0=0.003$ which we know is close to the upper limit of the Ohmic
regime for DC transport~\cite{aurora}.
Dependences of $\sigma$ on the field amplitude for different
frequencies are shown in the right panel. As it is seen, the
conductance grows with frequency, as well as with electric field
that qualitatively agrees with the experimental results shown in
Figs.~\ref{fig3}, \ref{fig4} (right panels).

\begin{figure}[h]
\begin{center}
\includegraphics[width=0.49\columnwidth]{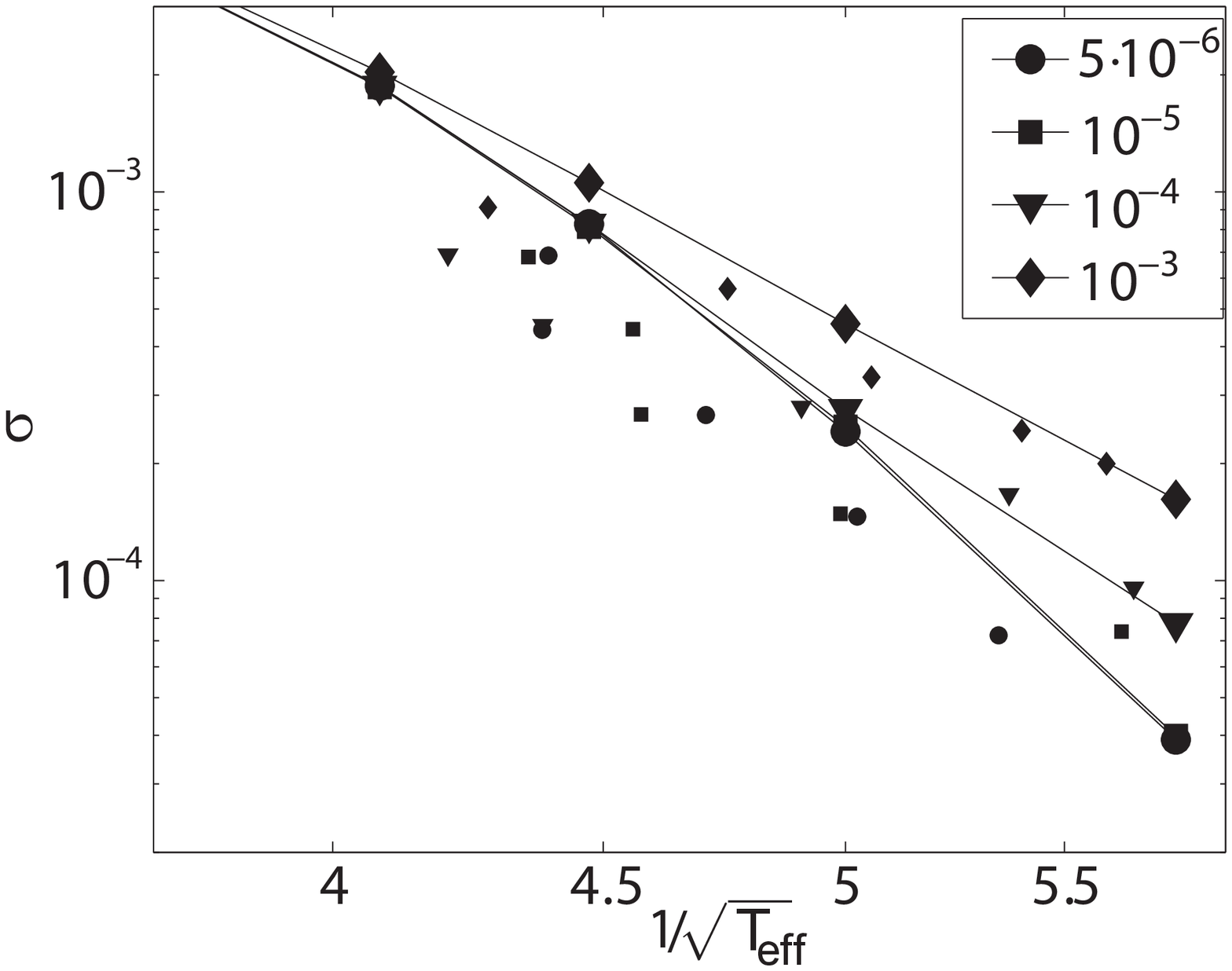} \hfill \includegraphics[width=0.49\columnwidth]{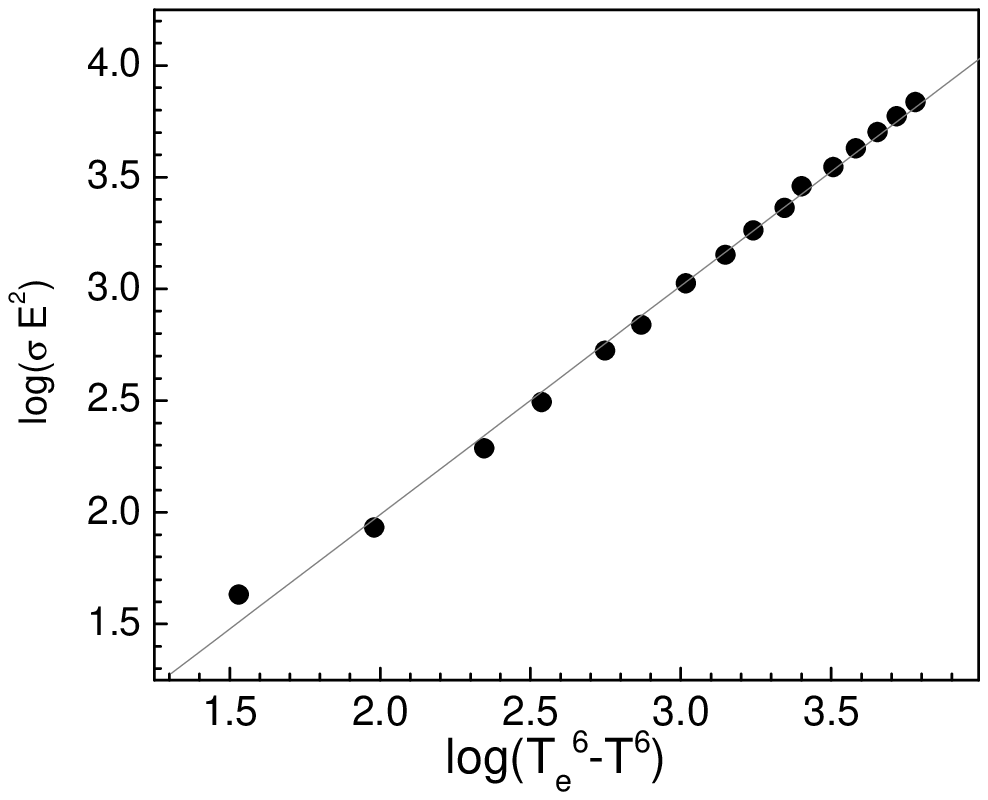}
\caption{\label{condVSTeff}
Left -- Conductance as a function of
  $1/\sqrt {T_\e}$ for various frequencies. The large symbols connected by
  lines show the ohmic conductivity (for which $T_\e=T$).  The
  small symbols show the non-Ohmic conductivity at a finite field. Each symbol
  corresponds to a different frequency. Right -- Relationship between $\sigma (T,E)E^2$ and $T_\e^6-T^6$ for sample 1 at $T=1.8$ K.
 }
\end{center}
\end{figure}

We can present the results
in a different form: assuming that the non-Ohmic behavior is due to
heating of the charge carriers we would expect that the conductance
at a
large field will be the same as the Ohmic conductance at a \textit{higher} temperature corresponding to the electron temperature. This is known to be approximately true for DC conductivity~\cite{caravaca,aurora}. Thus, we expect a relation of the form
\begin{equation}\label{eq:condVSTeff} \sigma(T_\e,E,\omega) =
\sigma (T,0,\omega)
\end{equation}
where $T_\e (E)$ is the effective electron temperature while $\sigma
(T,0,\omega)$ is the Ohmic AC conductance. In the simulations we can
check the accuracy of this relation by independent calculation of
$\sigma (\omega)$ and $T_\e$. The latter
was found
 from direct fitting of the Fermi function to the actual distribution function of
 occupied sites obtained by averaging over 20 states.
We also found the Ohmic conductivity as function of temperature for
different frequencies. If relation (\ref{eq:condVSTeff}) holds we
should find at a given frequency the following: as the electric
field is increased, both the effective temperature and the
conductivity increase in such a way that the points
$\sigma(T,E,\omega)$ follow the same curve as the Ohmic
conductivity. We plot $\sigma$ versus  $1/\sqrt{T_\e}$ for different
frequencies in Fig. \ref{condVSTeff} (left panel). As we can see,
relation (\ref{eq:condVSTeff}) is approximately satisfied.
%
Therefore, we have used it
as a \textit{definition} of the electron temperature
allowing us to extract it from experimental data on  $\sigma(T,E,\omega)$ and $\sigma (T,0,\omega)$.
Shown in the right panel of Fig.~\ref{condVSTeff} is the relationship between so-defined $T_\e$ and the
absorbed power, $\sigma (T,E,\omega) E^2$, for the lattice temperature of $1.8$ K.
Here we used averaged data for different frequencies since the frequency dependence is rather weak.
The results are compatible with the dependence $T_\e^6-T^6 \propto \sigma (T,E) E^2$. Interestingly, a
similar dependence was reported for amorphous InO$_x$ films, which are on the dielectric side of
a insulator-to-superconductor transition~\cite{ovadia}.  Though such a dependence was predicted for
a metallic state~\cite{schmid}, we are not aware of any
analytical  theory
explaining such a dependence in the insulating phase.

\paragraph{Conclusion}
Using
acoustic methods we have measured real part of nonlinear AC conductance in insulating 2D arrays of Ge-in-Si QDs.
The observed features 
-- inequality $\sigma_1 \gtrsim \sigma_2$, increasing
of $\sigma_1$ with the field amplitude,
its relatively weak frequency dependence, large energy exchange length $\mathcal{L} \gtrsim 1/ \sqrt{n}$
 -- essentially
\textit{contradict} to the predictions of the two-site model discussed in Sec.~\ref{introduction}.
The dissipation is rather due to electron transitions in multi-site clusters.
This is not surprising because the two-site model requires the close pairs to be isolated that
can be met only in dilute systems. The observed non-Ohmic behavior
is attributed to \textit{heating} of the charge carriers by AC
electrical field, as in the DC case~\cite{Gershenson2000}.
We have also performed 
simulations of AC conductance in a dense 2D lattice of localized states taking into account both distribution of their energies and Coulomb correlations. The results qualitatively agree with experiment confirming the above conclusion.
%

This work was supported by grant of RFBR 11-02-00223, grant of the
Presidium of the Russian Academy of Science, the Program
"Spintronika" of Branch of Physical Sciences of RAS.






\end{document}